\documentstyle[pra,aps,twocolumn,psfig,floats]{revtex}
\begin{document}
\draft

\title{Measurement of  quantum devices }

\author{Jarom\'{\i}r Fiur\'{a}\v{s}ek and Zden\v{e}k Hradil}

\address{Department of Optics, Palack\'{y} University, 17. listopadu 50,
772 07 Olomouc, Czech Republic}

\maketitle

\begin{abstract}
Maximum-likelihood estimation is applied  to identification  of an
unknown quantum mechanical process represented by a ``black box''.
In
contrast to linear reconstruction schemes the proposed approach
always yields physically sensible results. Its  feasibility is
demonstrated using the  Monte Carlo simulations for the two-level
system (single qubit).
\end{abstract}


During recent years great   attention has been devoted to the {\em
measurement of quantum state} of various simple quantum mechanical
systems. All proposed reconstruction techniques follow the common
underlying strategy: A set of measurements is performed on many
identically prepared copies of the quantum state which is then
estimated from the collected data. Feasible reconstruction schemes
were devised for a wide variety of systems including the modes of
running electromagnetic field (optical homodyne tomography
\cite{Smithey93,Vasilyev00} and unbalanced homodyning
\cite{Wallentowitz96}), cavity electromagnetic field
\cite{Lutterbach97,Bodendorf98}, motional state of ion in Paul trap
\cite{Wallentowitz95,Leibfried96}, vibrational state of the
molecule \cite{Dunn95} and spin \cite{Newton68}.

These significant achievements stimulated development of a new
remarkable branch of the reconstruction techniques that allow for
the experimental determination of the unknown {\em quantum
mechanical processes}
\cite{Poyatos97,Chuang97,DAriano98,Buzek98,Gutzeit00,Luis99b}. This
is of great practical importance, because such a technique may be
used, e.g. to  evaluate experimentally the performance of the
two-bit quantum gate -- a building block of quantum computers
\cite{Poyatos97}. The usual set-up considered also in this paper is
shown in Fig. 1. Note that a similar experimental configuration can
also allow for a complete characterization of quantum measurement
\cite{Luis99}. The input state prepared by an experimentalist and
characterized by a density matrix $\varrho_{\rm in}$ enters the
``black box'' where it is transformed into the output $\varrho_{\rm
out}$. The task for the experimentalist is to retrieve information
on the physical process hidden in the black box from the
measurements on the output states $\varrho_{\rm out}$. The only
assumption taken for granted here is that the mapping $\varrho_{\rm
out}={\cal{G}} \varrho_{\rm in}$ is {\em linear}, as dictated by
the linearity of quantum mechanics,
\begin{equation}
\varrho_{{\rm out},ij}=\sum_{kl} {\cal{G}}_{ij}^{kl} \varrho_{{\rm in},kl}.
\label{RHOINOUT}
\end{equation}
Here $\varrho_{ij}=\langle i|\varrho|j\rangle $ are density matrix
elements in some complete orthogonal basis of states spanning the
Hilbert space on which the density operator $\varrho$ acts. As
illustrated in Fig. 1, the system  may be entangled with the
environment and the transformation ${\cal{G}}$ need not preserve
purity of the state. The Green superoperator ${\cal{G}}$ can
describe a diverse variety of the physical processes, such as
unitary evolution, damping and decoherence. From the reconstructed
superoperator ${\cal{G}}$ one may further  estimate the Liouville
superoperator ${\cal L}$, which governs the evolution of density
matrix in the black box, $\dot{\varrho}={\cal{L}} \varrho$. If the
superoperator $\cal{L}$ exists, then
${\cal{G}}=\exp({\cal{L}}\tau)$, where $\tau$ is the interaction
time, and an inversion of this relation yields $\cal{L}$
\cite{DAriano98,Buzek98}.

The estimation of the elements ${\cal{G}}_{ij}^{kl}$
 by means of linear algorithms has been addressed in
several papers \cite{Poyatos97,Chuang97,DAriano98}. Provided that
mapping between
 known input and output states is given,  the unknown parameters
 ${\cal{G}}_{ij}^{kl}$ may be obtained from  (\ref{RHOINOUT}) using
 the system of linear equations. This linear reconstruction
procedure is simple and straightforward, but it suffers from one
significant drawback. The elements ${\cal{G}}_{ij}^{kl}$ are
estimated as a set of seemingly unrelated numbers. However,
${\cal{G}}_{ij}^{kl}$ cannot be arbitrary because the linear
mapping ${\cal{G}}$ must preserve the  positive semidefiniteness
and trace of the density matrix. These conditions  impose bounds on
the allowed values of ${\cal{G}}_{ij}^{kl}$. In this Rapid
Communication the superoperator  ${\cal{G}}$ is reconstructed using
maximum--likelihood (Max-Lik). It  allows  to incorporate naturally
 all the  constrains of quantum theory. Since one can only collect
a finite amount of data, the linear mapping  cannot be determined
exactly. In accordance with the probabilistic interpretation of the
quantum theory, the Max-Lik estimation answers the question {\em
``Which process is most likely to yield the measured data?''}.
However, Max--Lik solution is not only an estimation, but
represents a genuine quantum measurement associated with a quantum
device.

\begin{figure}[!h!]
\centerline{\psfig{file=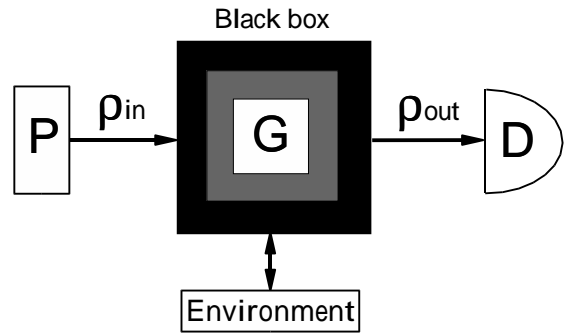,width=0.85\linewidth}}
\vspace*{4mm}
\caption{Sketch of experimental set-up for determination of the quantum-mechanical
process. The input state $\varrho_{\rm in}$ is prepared in the
preparator $P$ and enters the black box  where it is transformed to the
output state $\varrho_{\rm out}={\cal{G}}\varrho_{\rm in}$ which may be
entangled with the environment. The detector $D$
measures some observable of the output $\varrho_{\rm out}$.}
\end{figure}

\newpage

Due to its nonlinearity, the Max-Lik estimation is computationally
much more expensive task then the linear procedures. This is the
prize paid for the physically sound result. Max--Lik estimation has
been applied to various problems recently: To the measurements of
the quantum phase shift \cite{Hradil96}, a coupling constant
between atom and a cavity electromagnetic field \cite{Mabuchi96},
or the parameters of quantum-optical Hamiltonian \cite{DAriano00}.
Reconstruction of generic quantum state using the Max--Lik
estimation and its interpretation as quantum measurement has been
proposed
 in \cite{Hradil97}. Subsequent Monte Carlo simulations, performed
for the quantum states of electromagnetic field modes and spin
\cite{Banaszek98,Banaszek99,Hradil00}, illustrated a feasibility of
this technique. Here we shall demonstrate that the Max-Lik
estimation is also suitable for determination of the generic
quantum mechanical processes.

The sought superoperator ${\cal{G}}$ can be found as that one
maximizing the likelihood function ${\cal{L}}[{\cal{G}}]$. Let us
consider $n$ measurements described by positive operator-valued
measures (POVM) $\Pi^{(m)}$, $m=1,\ldots,n$. Then
${\cal{L}}[{\cal{G}}]$ reads
\begin{eqnarray}
{\cal{L}}[{\cal{G}}]&=&\prod_{m=1}^{n}\left( {\rm
Tr}\left[\Pi^{(m)}
\varrho_{\rm out}^{(m)}\right]\right)^{f_m} \nonumber \\
&=&\prod_{m=1}^{n} \left(\sum_{ijkl}
                 \Pi^{(m)}_{ij}{\cal{G}}_{ji}^{kl}
                 \varrho_{{\rm in},kl}^{(m)}\right)^{f_m},
\label{LR}
\end{eqnarray}
where $f_m$ is (relative) frequency for detection of $\Pi^{(m)}$.
Maximum of this function should be found in the domain of
physically allowed superoperators ${\cal{G}}$, whose determination
is crucial for successful implementation of the Max-Lik estimation.
The {\em linear positive map} (\ref{RHOINOUT}) can be conveniently
cast into the form which explicitly preserves the positive
semidefinitness of the density matrix  \cite{Chuang97},
\begin{equation}
\varrho_{\rm out}=\sum_i A_i \rho_i A_i^\dagger.
\label{RHOINOUTPOSITIVE}
\end{equation}
It follows from the condition ${\rm Tr}\, \varrho_{\rm out}=1$ that
\begin{equation}
\sum_i A_i^\dagger A_i =I,
\label{ACONSTRAINT}
\end{equation}
where $I$ denotes the identity operator. Further we can expand
$A_i$ in some complete operator basis $\tilde{A}_j$,
\begin{equation}
A_i=\sum_j  c_{ij} \tilde{A}_j.
\label{AEXPANSION}
\end{equation}
 If we deal with $N$ level system $|i\rangle$, $i=0,\ldots,N-1$, then
it is natural to choose the $N^2$ basis operators as
\begin{equation}
\tilde{A}_{Ni+j}=|i\rangle\langle j|, \qquad
i,j=0,\ldots,N-1,
\label{ABASIS}
\end{equation}
but other constructions are possible. Inserting (\ref{AEXPANSION})
into (\ref{RHOINOUTPOSITIVE}), we find that
\begin{equation}
\varrho_{\rm out}=\sum_{jk}\chi_{jk} \tilde{A}_j \varrho_{\rm in}
\tilde{A}_k^\dagger,
\label{RHOINOUTCHI}
\end{equation}
where
\begin{equation}
\chi_{jk}=\sum_i c_{ij} c_{ik}^\ast, \qquad j,k=0,\ldots, N^2-1.
\end{equation}
Thus $\chi$ is positive semidefinite hermitian matrix. This is the
desired condition revealing a domain of the allowed parameters
${\cal{G}}_{ij}^{kl}$ (or, alternatively, $\chi_{ij}$). The matrix
$\chi$ is parameterized by $N^4$ real numbers, but the condition
(\ref{ACONSTRAINT}) imposes $N^2$ real constraints so that the
number of independent parameters reads $N^4-N^2$. When the operator
expansion (\ref{AEXPANSION}) is substituted into Eq.
(\ref{ACONSTRAINT}), one obtains
\begin{equation}
\sum \chi_{jk} a_{mn}^{jk}=\delta_{mn}, \qquad m,n=0,\ldots,N-1,
\label{CHICONSTRAINT}
\end{equation}
where $a_{mn}^{jk}=\langle m|\tilde{A}_k^\dagger \tilde{A}_j
|n\rangle$. The constraints can be also expressed in terms of
${\cal{G}}_{ij}^{kl}$,
\begin{equation}
\sum_i {\cal{G}}_{ii}^{kl}=\delta_{kl}.
\label{RCONSTRAINT}
\end{equation}
From these $N^2$ linear constraints one can easily express $N^2$
real parameters in terms of the remaining $N^4-N^2$  ones and thus
achieve minimal parameterization.

The relation between $\chi$ and ${\cal{G}}$ can be found by comparing
Eqs. (\ref{RHOINOUT}) and (\ref{RHOINOUTCHI}),
\begin{equation}
{\cal{G}}_{ij}^{kl}=\sum_{m,n=0}^{N^2-1}
\langle i|\tilde{A}_m |k\rangle \langle l| A_n^\dagger |j\rangle \chi_{mn}.
\label{RCHI}
\end{equation}
This formula simplifies considerably if the basis (\ref{ABASIS}) is
chosen ${\cal{G}}_{ij}^{kl}=\chi_{iN+k,jN+l}$. To provide an
explicit example, let us consider a two-level system (single
qubit). The matrix $\chi$ can be  expressed in terms of
${\cal{G}}_{ij}^{kl}$ as follows,
\begin{equation}
\chi=\left(
\begin{array}{cccc}
{\cal{G}}_{00}^{00} & {\cal{G}}_{00}^{01} & {\cal{G}}_{01}^{00}
& {\cal{G}}_{01}^{01} \\[1.5mm]
{\cal{G}}_{00}^{10} & {\cal{G}}_{00}^{11} & {\cal{G}}_{01}^{10}
& {\cal{G}}_{01}^{11} \\[1.5mm]
{\cal{G}}_{10}^{00} & {\cal{G}}_{10}^{01} & {\cal{G}}_{11}^{00}
& {\cal{G}}_{11}^{01} \\[1.5mm]
{\cal{G}}_{10}^{10} & {\cal{G}}_{10}^{11} & {\cal{G}}_{11}^{10}
& {\cal{G}}_{11}^{11}
\end{array}
\right),
\label{CHIMATRIX}
\end{equation}
and the constraints (\ref{ACONSTRAINT}) yield
$
{\cal{G}}_{11}^{kl}=\delta_{kl}-{\cal{G}}_{00}^{kl}.
$
 Thus $\chi$ is parameterized by $16-4=12$ real parameters that can
be collected in a vector
\begin{eqnarray}
\vec{G}&=&
({\cal{G}}_{00}^{00},{\cal{G}}_{00}^{11}, {\rm
Re}\,{\cal{G}}_{00}^{01},{\rm Im}\,{\cal{G}}_{00}^{01}, {\rm
Re}\,{\cal{G}}_{01}^{00}, {\rm Im}\,{\cal{G}}_{01}^{00},
\nonumber \\&&
{\rm Re}\,{\cal{G}}_{01}^{10},{\rm Im}\,{\cal{G}}_{01}^{10}, {\rm
Re}\,{\cal{G}}_{01}^{01},{\rm Im}\,{\cal{G}}_{01}^{01}, {\rm
Re}\,{\cal{G}}_{01}^{11},{\rm Im}\,{\cal{G}}_{01}^{11}
 )
\end{eqnarray}
 Note that ${\cal{G}}_{ij}^{kl}=({\cal{G}}_{ji}^{lk})^{\ast}$ since
$\chi$ is hermitian. Additional constraints on
${\cal{G}}_{ij}^{kl}$ follow from the  positive semidefiniteness of
$\chi$. All four main subdeterminants of the matrix
(\ref{CHIMATRIX}) should be non-negative. This can be easily
checked for each  ${\cal{G}}$ where the likelihood function
(\ref{LR}) is evaluated. If (\ref{CHIMATRIX}) is not positive
semidefinite, then one may simply put ${\cal{L}}[{\cal{G}}]=0$. The
maximum of $\cal{L}$ can be  found for example  with the help of
downhill-simplex algorithm.  In case of 2 level system it is
sufficient to search for the maximum in the finite volume subspace
of 12 dimensional space.

Alternatively, one can find the maximum by setting to zero all
derivatives of  ${\cal{L}}[{\cal{G}}]$ with respect of
${\cal{G}}_{ij}^{kl}$. It is convenient to work with the
log-likelihood function. The constraints (\ref{CHICONSTRAINT}) must
be incorporated by introducing $N^2$ (complex) Lagrange multipliers
$\lambda_{mn}=\lambda_{nm}^{\ast}$. Thus one arrives at
\begin{equation}
\frac{\partial }{\partial {\cal{G}}_{ij}^{kl}}\left[
{\ln \cal{L}}[{\cal{G}}]
-\sum_{mn}\lambda_{mn}\sum_{p} {\cal{G}}_{pp}^{mn} \right]=0.
\label{RMAXIMUM}
\end{equation}
 Eqs. (\ref{RCONSTRAINT}) and (\ref{RMAXIMUM}) represent a system
of $N^4+N^2$ nonlinear equations which must be solved for $N^4$
elements ${\cal G}_{ij}^{kl}$ and $N^2$ Lagrange multipliers
$\lambda_{mn}$. On inserting the explicit expression for the
likelihood function (\ref{LR}) into Eq. (\ref{RMAXIMUM}) one
obtains,
\begin{equation}
\lambda_{kl}\delta_{ab}=\sum_{m}\frac{f_m}{p_{m}} \Pi_{ba}^{(m)}
\varrho_{{\rm in},kl }^{(m)},
\label{LMN}
\end{equation}
where we have introduced
\begin{equation}
p_m =  {\rm Tr} [\sum_i A_i \rho_{\rm in}^{(m)} A_i^{\dagger}
\Pi^{(m)} ]= {\rm Tr}\,\left(\Pi^{(m)}{\cal{G}}
\varrho_{\rm in}^{(m)}\right).
\end{equation}
As follows from  Eq. (\ref{LMN})  $\lambda$ is positive definite
hermitian matrix.

The extremal equation  may be rewritten to the form suitable for
iterative solution. Multiplying  eq. (\ref{LMN}) by
$(\lambda^{-1})_{ln} {\cal{G}}_{ac}^{kp} $ and summing over
$a,k,l$, one  gets
\begin{equation}
{\cal{G}}_{bc}^{np}=\sum_{m}\frac{f_m}{p_m}\sum_{a,k,l} \Pi_{ba}^{(m)} \,
\varrho_{\rm in,kl}^{(m)} \, (\lambda^{-1})_{ln} \, {\cal{G}}_{ac}^{kp}.
\label{RITER}
\end{equation}
Convenient form of  Lagrange multipliers $\lambda_{mn}$ may be
found by inserting  eq. (\ref{RITER}) into (\ref{RCONSTRAINT})
\begin{equation}
\lambda_{ij}=\sum_{m}\frac{f_m}{p_m}\sum_{a,k,p}
\Pi_{ka}^{(m)} \, {\cal{G}}_{ak}^{pi} \, \varrho_{{\rm in},pj}^{(m)}.
\label{LAMBDA}
\end{equation}
The system of nonlinear equations (\ref{RITER}) and (\ref{LAMBDA})
for the elements of ${\cal{G}}$ can be conveniently solved by repeated
iterations.

The theory may be formulated   in terms of the operators $A_i$,
$A_i^\dagger$.  It is helpful to define a hermitian operator
\begin{equation}
\lambda=\sum_{mn} \lambda_{mn} |m\rangle \langle n|.
\end{equation}
The maximum of log-likelihood function can be formally found as the
relation
\begin{equation}
\frac{\partial}{\partial A_i^\dagger}
\left(
\ln {\cal{L}}[\{ A_i \}]-{\rm Tr}\,[\lambda \sum_i A_i^\dagger A_i]
\right)=0.
\end{equation}
On performing the differentiation with respect to $A_i^\dagger$, and
solving for $A_i$, we obtain
\begin{equation}
A_i=\sum_m \frac{f_m}{p_m} \Pi^{(m)} A_i \varrho_{\rm in}^{(m)}
    \lambda^{-1}.
\label{AITER}
\end{equation}
Next we multiply (\ref{AITER}) from the left by operator $A_i^\dagger$
and sum over $i$. Taking into account the constraint
(\ref{ACONSTRAINT}), we find
\begin{equation}
\lambda =\sum_{m} \frac{f_m}{p_m}\sum_i
A_i^\dagger  \Pi^{(m)}  A_i  \varrho_{\rm in}^{(m)},
\label{LAMBDAO}
\end{equation}
which is equivalent to (\ref{LAMBDA}). Notice that
 ${\rm Tr} \lambda = \sum_m f_m =  1.$

The procedure of Max-Lik estimation may be interpreted as a
generalized measurement. To show this explicitly, let us put $k =
l$ in the relation (\ref{LMN}) and add all the elements over $k$
\begin{equation}
{\rm Tr}\, \lambda \, \delta_{ab} = \sum_{m} \frac{f_m}{p_m}
\Pi^{(m)}_{ba} {\rm  Tr}\, \varrho_{\rm in}^{(m)}.
\end{equation}
Since  all the traces are equal to 1, this relation reads in the
operator form
\begin{equation}
 \sum_{m} \frac{f_m}{p_m}
\Pi^{(m)} = I.
\label{closure}
\end{equation}
This is nothing else as the closure relation for renormalized
positive valued operator measures
$$ \Pi^{\prime (m)} = \frac{f_m}{p_m} \Pi^{(m)}.$$
 Moreover, in spite of the fact that the relation used by standard
reconstructions $p_m = f_m $ cannot be fulfilled in general, the
analogous relation for the renormalized POVM is identically true
\begin{equation}
p^{\prime }_{m} \equiv {\rm Tr} [\sum_i A_i \rho_{in}^{(m)}
A_i^{\dagger} \Pi^{\prime (m)} ] \equiv f_m.
\label{data}
\end{equation}
This indicates the privileged role of Max-Lik estimation in analogy
with the  quantum state estimation \cite{Hradil97}. Max--Lik
estimation  represents a genuine quantum measurement. Properties of
a quantum black box are determined using the closure relation
(\ref{closure}) for a POVM, expectation values of which are the
registered data (\ref{data}).

 In the rest of the paper we  demonstrate the feasibility of our
approach by means of Monte Carlo simulations for two-level system
(a single qubit). We shall consider spin $1/2$ system. The detector
$D$ shown in Fig. 1 is Stern-Gerlach apparatus measuring the spin
projections along one of three axes $x$, $y$, $z$. We further
assume that $\varrho_{\rm in}$ is prepared in one of six
eigenstates $|\uparrow_j\rangle$, $|\downarrow_j\rangle$ of the
spin projectors (Pauli matrices) $\sigma_j$, $j=x,y,z$,
$\sigma_j |\uparrow_j \rangle=|\uparrow_j \rangle,$
and \qquad
$\sigma_j |\downarrow_j \rangle=-|\downarrow_j \rangle$.
We choose the basis $|0\rangle=|\downarrow_z\rangle$ and
 $|1\rangle=|\uparrow_z\rangle$. Each of the six input states is
prepared $3{\cal N}$ times. At the output, one measures $\cal{N}$
times the spin along each of the three axes $x,y,z$. The
corresponding six projectors read
$
\Pi_j=|j\rangle \langle j|,$ $
 j=\uparrow_x,\downarrow_x, \uparrow_y,\downarrow_y,
\uparrow_z,\downarrow_z.
$
Let $f_{jk}$ denote the relative frequency of projections to the
state $|k\rangle$ measured for the input state $|j\rangle$.
 The likelihood function can be expressed as product of $36$ terms,
\begin{equation}
 {\cal{L}}[{\cal{G}}]=
\prod_{j,k}\Bigl(\langle k| \,{\cal{G}}\bigl[| j\rangle\langle j|\bigr]\,
|k\rangle\Bigr)^{f_{jk}},
\end{equation}
where $j,k\in\{\uparrow_x,\downarrow_x, \uparrow_y,\downarrow_y,
\uparrow_z,\downarrow_z\}$.

\begin{figure}[t]
\centerline{\psfig{figure=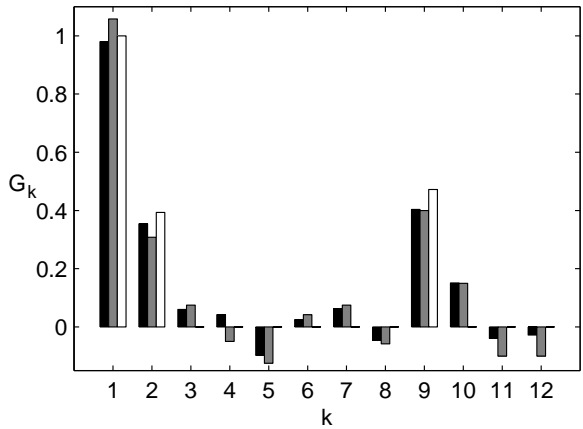,width=0.9\linewidth}}
\vspace*{2mm}
\caption{Reconstructed elements of the superoperator ${\cal{G}}$
 plotted in  the form of the vector $\vec{G}$. Bars correspond to
the  Max-Lik estimation (black), linear inversion (grey), and exact
values (hollow). Missing hollow bars indicate the zero true values.
The superoperator describes the process of  damping,
 $\Gamma_{||}=0.5$ and $\Gamma_{\perp}=0.75$, ${\cal{N}}=20$. }
\end{figure}

In our simulations, the black box of the  Fig. 1 corresponds to
the damping of $\rho_{\rm in}$,
\begin{equation}
\rho_{\rm out}=\left(
\begin{array}{ccc}
1-\rho_{{\rm in},11} e^{-\Gamma_{||}} & &
\rho_{{\rm in},01} e^{-\Gamma_{\perp}} \\[2mm]
\rho_{{\rm in},10} e^{-\Gamma_{\perp}} & &
\rho_{{\rm in},11} e^{-\Gamma_{||}}
\end{array}
\right).
\label{PROCESS}
\end{equation}
Here $2\Gamma_{\perp}\geq \Gamma_{||}\geq 0$ are transversal and
longitudinal decay parameters. The elements of reconstructed
superoperator are depicted in the Fig. 2. The solution was obtained
by iterations of eqs. (\ref{RITER}) and (\ref{LAMBDA}). For the
total amount of $360$ measurements the Max-Lik estimate (black) is
very close to the exact values ${\cal{G}}$ (hollow). Notice that
Max-Lik provides always physically sound result, on the contrary to
the linear inversion (grey).

Properties of transforming systems  are of interest in any physical
theory. The developed formalism shows how to describe it as a
genuine quantum measurement. Quantum systems consisting of  spins,
two entangled  or three entangled  (GHZ) qubits are tractable due
to their low dimensionality. Proper and full quantum description of
possible transformations of such systems is, however, more
advanced, since it is characterized by 12, 240, or even 4032
parameters.

This work was  supported by Grant  LN00A015 of the Czech Ministry
of Education. This paper is dedicated to the anniversary of 65th
birthday of Prof. Jan Pe\v{r}ina.


\begin{references}

\bibitem{Smithey93}
D.T. Smithey, M. Beck, M.G. Raymer and A. Faridani,
Phys. Rev. Lett. {\bf 70}, 1244 (1993);
S. Schiller, G. Breitenbach, S.F. Pereira, T. M\"{u}ller, and J. Mlynek,
Phys. Rev. Lett. {\bf 77}, 2933 (1996);

\bibitem{Vasilyev00}
M. Vasilyev, S-K Choi, P. Kumar, and G. M. D'Ariano,
Phys. Rev. Lett. {\bf 84}, 2354 (2000).


\bibitem{Wallentowitz96}
S. Wallentowitz and W. Vogel,
Phys. Rev. A {\bf 53}, 4528 (1996);
K. Banaszek and K. W\'{o}dkiewicz,
Phys. Rev. Lett. {\bf 76}, 4344 (1996).


\bibitem{Lutterbach97}
L.G. Lutterbachand L. Davidovich,
Phys. Rev. Lett. {\bf 78}, 2547 (1997).


\bibitem{Bodendorf98}
C.T. Bodendorf, G. Antesberger, M.S. Kim, and H. Walther,
Phys. Rev. A {\bf 57}, 1371 (1998).


\bibitem{Wallentowitz95}
S. Wallentowitz and W. Vogel,
Phys. Rev. Lett. {\bf 75}, 2932 (1995).


\bibitem{Leibfried96}
D. Leibfried,
D.M. Meekhof, B.E. King, C. Monroe, W.M. Itano, and D.J. Wineland,
Phys. Rev. Lett. {\bf 77}, 4281 (1996).

\bibitem{Dunn95}
T.J. Dunn, I. A. Walmsley, and S. Mukamel,
Phys. Rev. Lett. {\bf 74}, 884 (1995);
C. Leichtle, W.P. Schleich, I.Sh. Averbukh, and M. Shapiro,
Phys. Rev. Lett. {\bf 80}, 1418 (1998).

\bibitem{Newton68}
R.G. Newton and B. Young,
Ann. Phys. {\em New York} {\bf 49}, 393 (1968).




\bibitem{Poyatos97}
J. F. Poyatos, J. I. Cirac, and P. Zoller,
Phys. Rev. Lett. {\bf 78}, 390 (1997).


\bibitem{Chuang97}
I.L. Chuang and  M.A. Nielsen, J. Mod. Opt. {\bf 44}, 2455 (1997).


\bibitem{DAriano98}
G. M. D'Ariano and L. Maccone,
Phys. Rev. Lett. {\bf 80}, 5465 (1998);
Fortschr. Phys. {\bf 46}, 837 (1998).

\bibitem{Buzek98}
V. Bu\v{z}ek, Phys. Rev. A {\bf 58}, 1723 (1998).

\bibitem{Gutzeit00}
R. Gutzeit, S. Wallentowitz, and W. Vogel, Phys. Rev. A {\bf 61}
062105 (2000).

\bibitem{Luis99b}
A. Luis and L.L. S\'{a}nchez-Soto,
Phys. Lett. A {\bf 261}, 12 (1999).



\bibitem{Luis99}
A. Luis and L.L. S\'{a}nchez-Soto,
Phys. Rev. Lett. {\bf 83}, 3573 (1999);

\bibitem{Hradil96}
Z. Hradil, R. My\v{s}ka, J. Pe\v{r}ina, M. Zawisky, Y. Hasegawa, and H.
Rauch, Phys. Rev. Lett. {\bf 76}, 4295 (1996).

\bibitem{Mabuchi96}
H. Mabuchi, Quantum semiclass. Opt. {\bf 8}, 1103 (1996).


\bibitem{DAriano00}
G. M. D'Ariano, M.G.A. Paris, and M.F Sacchi,
Phys. Rev. A {\bf 62}, 023815 (2000).



\bibitem{Hradil97}
Z. Hradil, Phys. Rev. A {\bf 55}, R1561 (1997);
Z. Hradil, J. Summhammer, and H. Rauch, Phys. Lett. A {\bf 261},
20 (1999).

\bibitem{Banaszek98}
K. Banaszek, Phys. Rev. A {\bf 57}, 5013 (1998).

\bibitem{Banaszek99}
K. Banaszek, G.M. D'Ariano, M.G.A. Paris, and M. F. Sacchi,
Phys. Rev. A {\bf 61}, 010304(R) (1999).

\bibitem{Hradil00}
Z. Hradil, J. Summhammer, G. Badurek, and H. Rauch,
Phys. Rev. A {\bf 62}, 014101 (2000).


\end{references}
\end{document}